\documentclass[pra,aps,12pt,onecolumn,superscriptaddress,showpacs,floats,citeautoscript,floatfix,footinbib]{revtex4-1}

\usepackage[latin1]{inputenc}
\usepackage[intlimits,tbtags]{amsmath}
\usepackage{color}
\usepackage{amsfonts}
\usepackage{amssymb}
\usepackage{graphicx}
\usepackage{bm}
\usepackage{dsfont}

\newcommand{\bra}[1]{\ensuremath{\langle #1|}}
\newcommand{\ket}[1]{\ensuremath{|#1\rangle}}

\newcommand{\ie}{\textit{i.e. }}
\newcommand{\eg}{\textit{e.g. }}
\newcommand{\real}{\ensuremath{\mathfrak{Re}}}
\newcommand{\imag}{\ensuremath{\mathfrak{Im}}}

\begin{document}

\title{Optimal Control of the Electronic Current Density -- An application to one- and two-dimensional one-electron systems}

\author{David Kammerlander}\email{david.kammerlander@univ-lyon1.fr}
\affiliation{Laboratoire de Physique de la Mati\`ere Condens\'ee et Nanostructures (LPMCN), Universit\'e Claude Bernard Lyon 1 and CNRS, 69100 Villeurbanne, France }
\author{Alberto Castro}
\affiliation{Institute for Biocomputation and Physics of Complex Systems (BIFI), University of Zaragoza, E-50018 Zaragoza, Spain}
\author{Miguel A. L. Marques}
\affiliation{Laboratoire de Physique de la Mati\`ere Condens\'ee et Nanostructures (LPMCN), Universit\'e Claude Bernard Lyon 1 and CNRS, 69100 Villeurbanne, France }

\date{\today}

\begin{abstract}
  Quantum optimal control theory is a powerful tool for engineering
  quantum systems subject to external fields such as the ones created
  by intense lasers. The formulation relies on a suitable definition
  for a target functional, that translates the intended physical objective 
  to a mathematical form. We propose the use of target
  functionals defined in terms of the one-particle density and its
  current. A strong motivation for this is the possibility of using
  time-dependent density-functional theory for the description of the
  system dynamics. We exemplify this idea by defining an objective
  functional that on one hand attempts a large overlap with a target
  density and on the other hand minimizes the current. The latter
  requirement leads to optimized states with increased stability,
  which we prove with a few examples of one- and two-dimensional
  one-electron systems.
\end{abstract}

\maketitle

\section{Introduction}
Quantum control~\cite{brif:10} is concerned with the detailed
manipulation of systems in the microscopic world. Its most common
application is the design of laser fields capable of triggering a
given response of some target system. Under this umbrella name of
quantum control we may include both experimental techniques, the
theoretical description of those, and the mathematical techniques used
to predict the controlling external fields. Experimentally, the field
has advanced very rapidly in the last years due to the convergence of
several important paths: most notably, the development of ultra-fast
and ultra-intense laser sources, the appearance of pulse
shapers~\cite{brixner:04,shverdin:05,parker:08,nuernberger:07,znakovskaya:09,krausz:09},
and the invention of the adaptive feedback technique~\cite{judson:92,assion:98,meshulach:98,weinacht:99,bartels:00}.

Theoretically, the most general framework used to address quantum
control problems is quantum optimal control theory
(QOCT)~\cite{werschnik:07,shi:88,peirce:88,kosloff:89}. It is
formulated as the problem of maximizing a functional that
quantifies to what extent the given objective is met by a certain laser
field. The result is typically an algorithm that requires, as main
steps, the propagation of the time dependent Schr{\"{o}}dinger
equation that describes the evolution of the system (and, usually, of a
similar equation that is propagated backwards). The ab initio solution
of these equations is, unfortunately, prohibitively expensive for most
systems due to the complexity of their wavefunctions.

Time-dependent density-functional theory
(TDDFT)~\cite{runge:84,marques:06} provides in many cases a viable
solution for this problem. Recently, the merger of QOCT and TDDFT has
been proposed and numerically demonstrated~\cite{castro:10}. Within
TDDFT the electron density, that can be obtained from the solution of
the so-called time dependent Kohn-Sham (KS) equations, substitutes the
many-electron wavefunction as the main object to be manipulated. 

The targets of QOCT are usually defined as functionals of
many-electron wavefunctions. For example, if one wishes to increase
the population of a given excited state, the target is defined in
terms of the projector onto that particular excited state wave
function. However, these wavefunctions are in principle unavailable 
within the TDDFT framework that provides the electron density only~\footnote{
It is true that the solutions of the time dependent KS equations are 
wavefunctions, but since the KS system is a system of non-interacting 
electrons, KS wavefunctions and wavefunctions of the real system are not identical; 
nonetheless, the respective densities are.}.
Therefore, if one is to use QOCT with TDDFT consistently, the targets 
should be defined as functionals of this electron density alone.

Yet, to construct better target functionals (whether or not TDDFT is used),
it may be useful to add an extra ingredient to the electron density, 
namely the electron current density. Note that the longitudinal
component of this current vector field is linked directly to the
density through the continuity equation, and can therefore be
obtained with TDDFT (the perpendicular component of the KS current
density is also conjectured to coincide with the many-electron one).

For example, it is easy to think of a density-functional target
constructed to maximize charge transfer between two regions of a
system. However, once the controlling field is switched off, even if
the charge has been effectively transferred, the final state may not
be stable. The inclusion of the current in the definition of the
target functional can be used, as will be demonstrated below, to
stabilize the final state of the propagation: if the current (or its
divergence) is small, the continuity equation ensures that the
time-variation of the density is also small. 

In order to demonstrate the usefulness and feasibility of including
the electron current density in the definition of target functionals,
we studied several model one-electron systems in one and two
dimensions. The point is to show that with this combined target one can obtain 
solutions that (i) are stabler if the current is minimized and that (ii) 
resemble results as if the target was 
formulated in terms of the wavefunction~\cite{zhu:98b}. 
Finally, we also note that we do not put forward the use of
current TDDFT~\cite{vignale:88}, but rather the inclusion of the
current density in the definition of the control targets. The proposed
equations have been implemented into the {\tt octopus}
code~\cite{marques:03,octopus:06}.

The paper is organized as follows: in Section~\ref{sec:theory} we
briefly outline the optimization scheme, and introduce the new control
functionals defined in terms of the electron density and its
current. In Section~\ref{sec:applications} we present numerical
examples that serve as a proof of principle. The conclusions are given
in Section~\ref{sec:conclusion}. Atomic units (a.u.) are used throughout.

\section{Formalism}
\label{sec:theory}

We consider a one-electron system governed at rest by the Hamiltonian
$H_0$ that interacts with a time-dependent control field $\epsilon(t)$
in the time span $t\in[0,T]$, such that the total Hamiltonian is given
by:
\begin{align}\label{eq:hamiltonian}
 H(t) &= H_0 - \mu \epsilon(t),&& 0 \leq t \leq T,
\end{align}
where $\mu$ is the dipole operator that couples with the electric
field $\epsilon(t)$. Without loss of generality, we will think of
$\epsilon(t)$ as the electric field of a laser pulse.

In QOCT one tries to find the control field that best realizes an
objective that in general is mathematically encoded into a functional
of the field $G[\epsilon]$. This is defined in the following way: the
field determines the evolution of the system, $\epsilon \to
\Psi[\epsilon]$, and this in turn determines $G$ through an intermediate functional $F$:
\begin{equation}
G[\epsilon] = F[\psi[\epsilon],\epsilon]\,.
\end{equation}
The functional $F$ must be carefully defined in order to ensure the fulfillment of
the objective. Typically, it is split into two pieces:
\begin{equation}
F[\psi,\epsilon] = \mathcal{O}[\psi] + \mathcal{F}[\epsilon]\,,
\end{equation}
where $\mathcal{O}[\psi]$ is the quantity that we really want to
optimize, whereas $\mathcal{F}$ imposes a penalty on undesired
features of the controlling fields: \eg high frequency components,
unrealistically high intensities, etc. In our case we have used the
following expression:
\begin{align}\label{eq:fluence}
 \mathcal{F}[\epsilon]= - \int_0^T dt \, \alpha (t) \epsilon^2(t).
\end{align}
If $\alpha=1$, $\mathcal{F}$ would be minus the \emph{fluence}, or
integrated energy of the laser pulse. Therefore, this term penalizes
pulses with too high intensities. The function $\alpha$ may then be
chosen to ensure that the laser pulse is smoothly switched on and off,
by penalizing non-zero field values near the beginning and the end of
the pulse:
\begin{align}\label{eq:penalty}
 \alpha(t)=\frac{1}{2} \left\{ {\rm erf} \bigg[t-\frac{T}{20} \bigg]- {\rm erf} \bigg[t-T+\frac{T}{20}\bigg]\right\}^{-1}\,,
\end{align}
where ${\rm erf}$ is the error function.

The search for the maxima of $G$ is substituted by the search for the
maxima of $F$, which however cannot be unconstrained, since the
evolution of the system must obey Schr{\"{o}}dinger's equation. The
formalism takes care of this by introducing a Lagrange functional:
\begin{align}\label{eq:lagrange_cond}
 \mathcal{L}[\psi, \chi, \epsilon]=- 2 \real\bigg [\int_0^T \bra{\chi(t)} \frac{\partial }{\partial t} + i H_0 -i \mu \epsilon(t) \ket{\psi(t)} \bigg],
\end{align}
where the auxiliary state $\chi$ is introduced as a Lagrange multiplier.
The functional whose critical points are to be found is:
\begin{align}
 \mathcal{J}[\psi,\chi,\epsilon]= \mathcal{O}[\psi] + \mathcal{F}[\epsilon] + \mathcal{L}[\psi, \chi, \epsilon].
\end{align}
The corresponding Euler-Lagrange equations are:
\begin{subequations}
\begin{align}
\label{eq:schrodinger_psi}
i \frac{\partial}{\partial t} \psi(t) &= \left[ H_0 - \mu\epsilon(t)\right] \psi(t)\,,
\\
\label{eq:schrodinger_psi_0}
\psi(0) & = \psi_0,
\\
\label{eq:schrodinger_chi}
i \frac{\partial}{\partial t} \chi(t) &= \left[ H_0 - \mu\epsilon(t)\right] \chi(t)\,, 
\\
\label{eq:schrodinger_chi_0}
\chi(\bm{r},T) & = \frac{\delta\mathcal{O}}{\delta \psi^*(\bm{r},T)}\,,
\\
\label{eq:control_field}
 \alpha(t) \epsilon(t) & = -\imag \, \bra{\chi(t)}\mu \ket{\psi(t)}.
\end{align}
\end{subequations}
Equation~\eqref{eq:schrodinger_psi} is merely Schr{\"{o}}dinger's
equation for the system $\psi$ under the influence of the controlling
laser $\epsilon$. It must be solved by \emph{forward} propagation,
since it is an initial value problem with boundary condition given by
Eq.~\eqref{eq:schrodinger_psi_0}. Equation~\eqref{eq:schrodinger_chi} is
also a Schr{\"{o}}dinger-like equation for the Lagrange multiplier
state $\chi$. However, in this case it must be propagated
\emph{backwards} since the boundary condition is given at the final
propagation time $T$. Finally, Eq.~\eqref{eq:control_field} couples
both $\psi$ and $\chi$, and permits to obtain the solution laser field
$\epsilon(t)$ (the dipole matrix element between states $\chi$ and
$\psi$ is the principal ingredient).

These equations are coupled, and their self-consistent solution
provides the critical points of the functional $G$. In order to find
one solution that corresponds to a maximum, we have in this work
utilized the monotonic algorithm described in
Ref.~\onlinecite{zhu:98}. It consists of a series of back- and forward
propagations.  For other possible algorithms to solve
Eqs.~\eqref{eq:schrodinger_psi} - \eqref{eq:control_field} we refer
the reader, for example, to Refs.~\onlinecite{maday:03,castro:09}.

Now we focus our attention to the precise form of the functional
$\mathcal{O}[\psi]$ (generally speaking, it is a functional of the
full evolution of the wave function $\psi$; however in the previous
equations we have assumed it depends only on the final state
$\psi(T)$). As discussed in the introduction above, we will assume
that it is a functional of the density $n$ and of the current density
$\bm{j}$: 
\begin{equation}
\label{eq:o}
\mathcal{O}[\psi] = O[n[\psi(T)],\bm{j}[\psi(T)]]\,,
\end{equation}
where $n$ and $\bm{j}$, for the one-electron case, read:
\begin{subequations}
\begin{align}
n[\psi](\bm{r}) & = \psi^*(\bm{r})\psi(\bm{r})\,,
\\
\bm{j}[\psi](\bm{r}) & = \imag \, [\psi^*(\bm{r}) \bm{\nabla} \psi \, (\bm{r})]\,.
\end{align}
\end{subequations}
The particular form for $\mathcal{O}$ given in Eq.~\eqref{eq:o} leads to
the following expression for the final-value condition for the
auxiliary wave function $\chi$ [Eq.~\eqref{eq:schrodinger_chi_0}]:

\begin{multline}
\label{eq:bondary-cond}
\frac{\delta\mathcal{O}}{\delta \psi^*(\bm{r},T)} = 
\psi(\bm{r},T)\frac{\delta O}{\delta n(\bm{r},T)}
-i \left[
  \bm{\nabla}\psi(\bm{r},T)
+ \frac{1}{2} \psi(\bm{r},T)\bm{\nabla}
\right]\cdot \frac{\delta O}{\delta \bm{j}(\bm{r},T)} \,,
\end{multline}

This expression is valid for any target functional defined in terms of
the density and its current. Let us narrow it down for a particular
case: we wish to maximize the overlap of the density $n$
with a target density $n_{\text{tg}}$ (at the end of the propagation).
We may define for that purpose a first functional $O_1$:
\begin{align}
  \label{eq:density_func}
 O_1[n] = - \int\!\! d^3r \,\left[ \sqrt{n(\bm{r}) \vphantom{ n_{\text{tg}} } }-\smash{ \sqrt{n_{\text{tg}}(\bm{r})} }\right]^2. 
\end{align}
It ranges in the interval $[-2,0]$ with its maximum at complete
overlap. However, if we only use this functional, the resulting
density at the end of the propagation will not be stable and change rapidly
after the controlling field is switched off. In order to stabilize the
achieved state, we may add a current dependent functional, making use
of the fact that the temporal behavior of the density is connected
with the current $\bm{j}(\bm{r},t)$ by the continuity equation:
\begin{align}\label{eq:continuity_eq}
\frac{\partial n(\bm{r},t)}{\partial t} = -\bm{\nabla j}(\bm{r},t)\,.
\end{align}
For any eigenstate of $H_0$ the divergence of the current vanishes,
and the density remains constant. Suppressing the current will assure a
stationary density, at least right after the controlling pulse is
switched off, and perhaps will ensure small variations
thereafter. Therefore we implemented an objective functional whose
maximum is at zero current, namely:
\begin{align}\label{eq:current_func}
 O_2[\bm{j}]=-w_c \int\!\! d^3r \, |\bm{j}(\bm{r})|^2.
\end{align}
Here, $w_c \ge 0$ is weighting the importance of the current suppression,
which is useful if one combines the objectives in
Eqs. \eqref{eq:density_func} and \eqref{eq:current_func}:
\begin{align}
\label{eq:density_current_functional}
\nonumber
\mathcal{O}[\psi] & = O[n[\psi(T)],\bm{j}[\psi(T)]] 
\\ & = 
O_1[n[\psi(T)]] + O_2[\bm{j}[\psi(T)]]\,.
\end{align}
In this case $w_c$ should also homogenize the dimensions of the
functionals. It only remains to rewrite Eq.~\eqref{eq:bondary-cond} for
this particular case:

\begin{multline}
\frac{\delta\mathcal{O}}{\delta \psi^*(\bm{r},T)} =
\psi(\bm{r},T) \sqrt{\frac{n_{\text{tg}}(\bm{r})}{n(\bm{r},T)}}
+ 2 i w_c \bm{j}(\bm{r},T) \cdot \bm{\nabla} \psi(\bm{r},T)
+ i w_c \psi(\bm{r},T) \bm{\nabla \cdot j}(\bm{r},T)\,.
\end{multline}

Two remarks are in order before moving on to applications: First,
instead of minimizing the current, we also applied this scheme to
minimizing the divergence of the current, since this is the quantity
to which the temporal behavior of $n$ is directly related by
Eq.~\eqref{eq:continuity_eq}. However, the more stringent minimization
of the total current turned out to be numerically preferable.
Finally, note that in the one-particle case, the wave function $\psi$
can be written as $\psi(\bm{r},t) =
\sqrt{n(\bm{r},t)}e^{iS(\bm{r},t)}$, and then the current density is
given by $\bm{j}(\bm{r},t)=n(\bm{r},t)\bm{\nabla}S(\bm{r},t)$. Hence
controlling the density and the current density amounts to controlling
the full wave function directly, since both objects contain the same
information.

\section{Applications}
\label{sec:applications}

In the following, we present three illustrative examples of the method
described above: electron transfer in both a 1D and a 2D asymmetric
double well, and the $1s \to 2p_x$ transition in a 2D model of the
hydrogen atom. In all calculations, there are a few parameters and
settings whose values have an influence in the final outcome, most
notably the value of the weight $w_c$ and the initial guess for the
laser field. Regarding the former, we observed that a value of around
$w_c=10$ led to reasonable convergence and stable results. Regarding
the initial guess, we found it advisable to do several calculations
with different starting points, since the search space contains many
local minima.

\subsection{1D asymmetric quantum well}

\begin{figure}
 \begin{center}
  \includegraphics[width=0.99\columnwidth,clip]{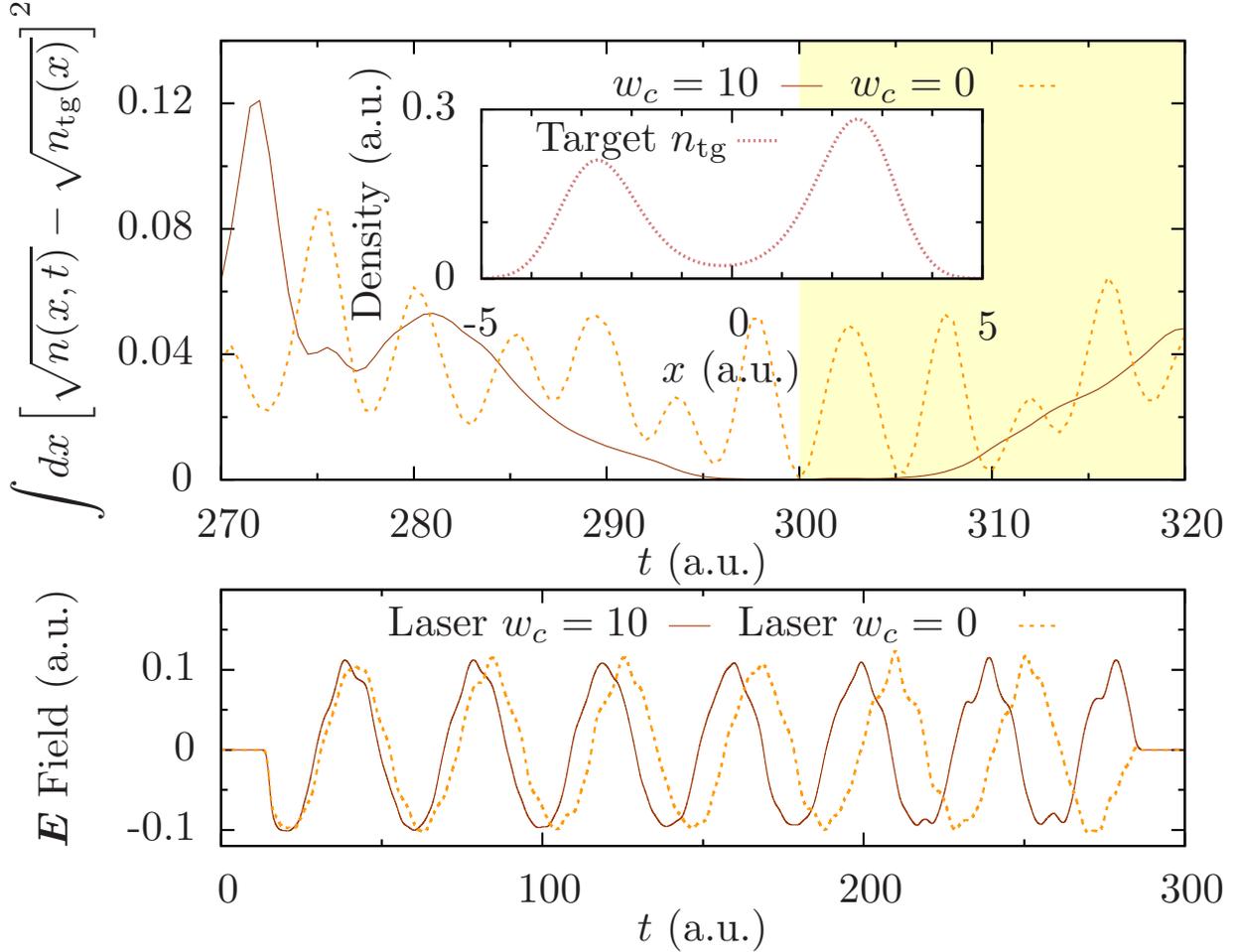} 
  \caption{\label{fig:1d_dens_diff}
    (Color online) 1D QW. The overlap between the controlled density $n$ and the target density $n_{\text{tg}}$ is shown in the top panel.
    For minimized current ($w_c=10$) the overlap with the target (found in the inset) at terminal time $T=300$~a.u. is better. Additionally,
    density $n$ is significantly more stable than without current control ($w_c=0$). The respective laser pulses are reported in the lower panel.} 
 \end{center}
\end{figure}

We consider a 1D asymmetric double well, defined by the following
potential function:
\begin{align}\label{eq:1dqw_potential}
 V(x)=\frac{1}{64}x^4 +\frac{1}{256}x^3 - \frac{1}{4} x^2\,,
\end{align}
which is sometimes used in quantum chemistry to model isomerisation
processes~\cite{doslic:98}. Note that even if we are usually referring
to electronic processes and electronic densities and currents, it need
not be the case, and the wave function may be that of a nuclear wave
packet. The goal is to guide a transition from the ground state (GS)
to a state whose density is equal to that of a superposition
of the GS and the first excited state (ES), within a total time of
$T=300$ a.u. We choose the superposition,
\begin{align}\label{eq:1dqw_superpos}
\psi_{\text{tg}}(x)= \frac{1}{\sqrt{2}} ( \psi_{\text{GS}}(x)+\psi_{\text{ES}}(x) ),
\end{align}
whose corresponding density $n_{\text{tg}} = |\psi_{\text{tg}}|^2$ is 
illustrated in the inset of the top panel of Fig.~\ref{fig:1d_dens_diff}.

As the target density $n_{\text{tg}} $ does not correspond to that of
a stationary state, it is crucial not only to maximize the overlap
according to Eq.~\eqref{eq:density_func}, but also to minimize the
current by setting $w_c=10$ in Eq.~\eqref{eq:current_func}.
The top panel of Fig.~\ref{fig:1d_dens_diff} shows for the cases of $w_c=10$ and
$w_c=0$ that a very good overlap is found at final time $T=300$~a.u. In
fact, mapping the target functional values on the interval $[0,1]$, we
obtain $0.99998$ for the first and $0.99989$ for the second case.
However, minimizing the current guarantees a significantly longer
lifetime of the density in its target shape. The two respective laser
fields are found in the bottom panel.

\subsection{2D asymmetric quantum well}

\begin{figure}
 \begin{center}
  \includegraphics[width=0.99\columnwidth,clip]{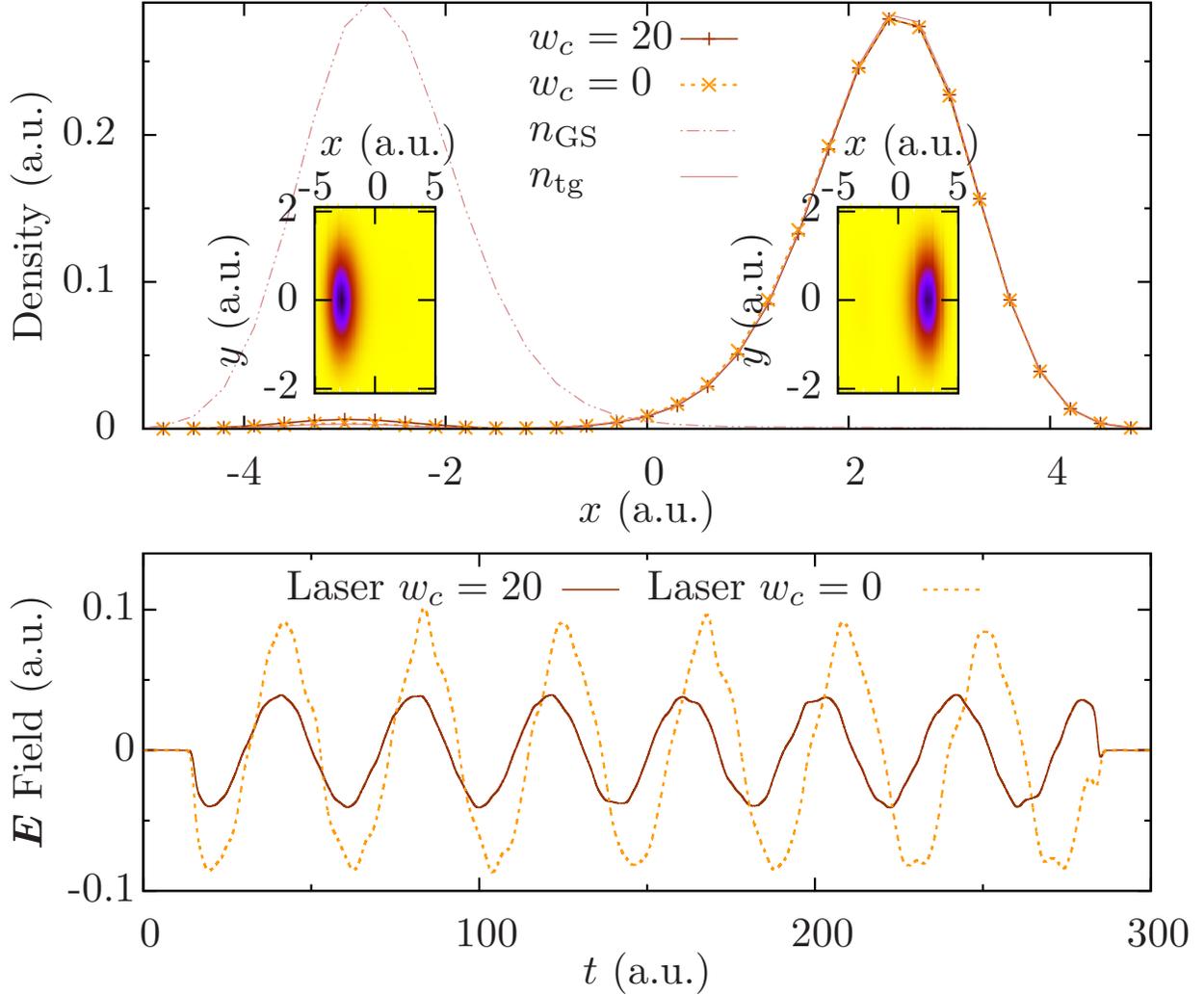} 
  \caption{(Color online) 2D QW. The top panel shows the optimized densities (with and
    without current suppression) as well as the densities of
    the GS and of the target $n_{\text{tg}}$ along the $x$-axis ($y=0$~a.u.). Both optimizations achieve
    an excellent target density overlap. The insets
    show the respective 2D plots of the GS and ES density. The bottom panel shows the
    optimal lasers.
  } \label{fig:2d_quwell_final_position}
 \end{center}
\end{figure}

A 2D asymmetric double quantum well can be realized by adding a
parabola in the $y$-direction to the previous 1D potential:
\begin{align}\label{eq:2dqw_potential}
 V(x,y)=\frac{1}{64}x^4 +\frac{1}{256}x^3 - \frac{1}{4} x^2 + \frac{1}{2}y^2\,.
\end{align}
The top panel of Fig.~\ref{fig:2d_quwell_final_position} displays the
densities of the GS and the first ES (indicated as $n_{\text{tg}}$), both as 1D plots along the $x$-axis ($y=0$) and as
density plots in the $xy$-plane (in the insets). The objective of this example is to
perform a density transfer from the GS to the ES with and
without suppressing the current, \ie with $w_c=20$ and $w_c=0$,
respectively.

\begin{figure}
 \begin{center}
  \includegraphics[width=0.99\columnwidth,clip]{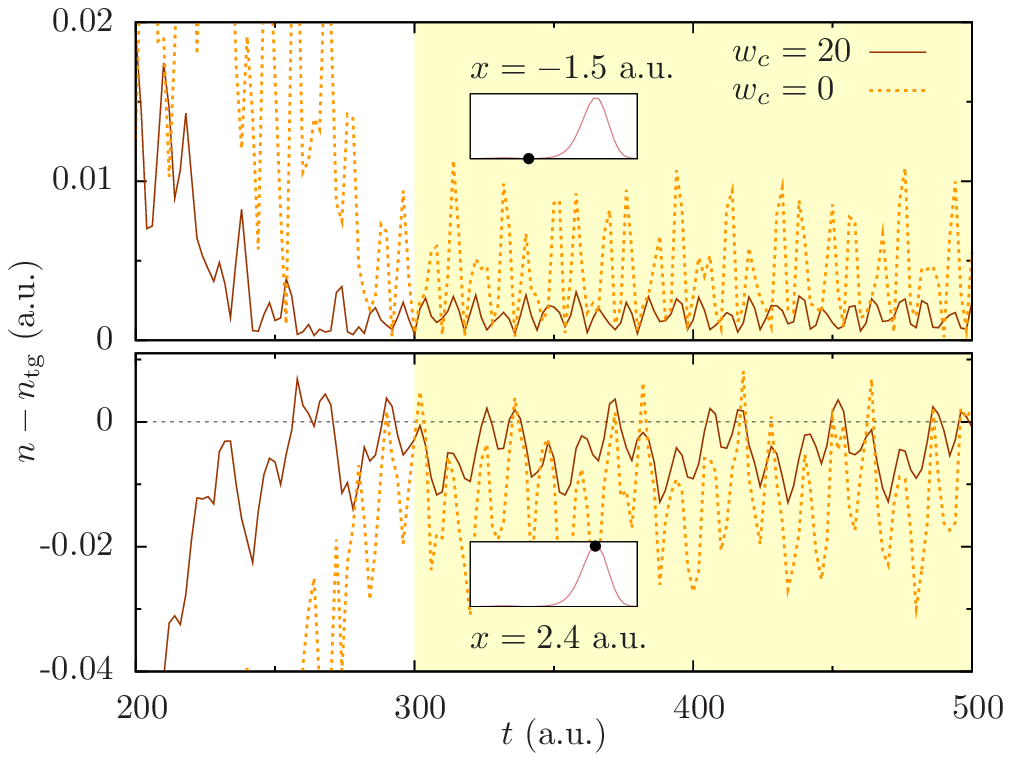} 
  \caption{(Color online) 2D QW. The difference between the controlled
    density $n$ and the target density $n_{\text{tg}}$ as a function
    of time at the node ($x=-1.5$~a.u., top panel) and at the maximum
    ($x=2.4$~a.u., bottom panel) of the target. The colored background
    marks the time $t \ge 300$~a.u. without laser. The insets in both panels are guides to the eye. They
    show the target density with a mark on the node and on the maximum, respectively. } 
    \label{fig:2d_quwell_differences}
 \end{center}
\end{figure}

As can be see in the top panel of
Fig.~\ref{fig:2d_quwell_final_position} both optimizations lead to a
very good overlap with the target density $n_{\text{tg}}$ at $T=300$~a.u. However, the
intensities of the optimizing laser fields were rather different: in
the case where the minimization of the current was enforced, the
electric field was roughly half the size of its counterpart without
current control. Also, the behavior of the respective densities
diverges considerably once the laser is switched off. This is
illustrated in Fig.~\ref{fig:2d_quwell_differences} where we show
the difference between the target $n_{\text{tg}}$ and the controlled
density $n$ at two crucial points along the $x$-axis. While the top
panel does so for the node of $n_{\text{tg}}$ at $x=-1.5$~a.u., the bottom
panel refers to its maximum at $x=2.4$~a.u. The requisite of 
current suppression induces much smaller oscillations of the
controlled density around a value closer to the target.


\subsection{2D hydrogen: transition $1s \rightarrow 2p_x$}

\begin{figure}
 \begin{center}
  \includegraphics[width=0.99\columnwidth,clip]{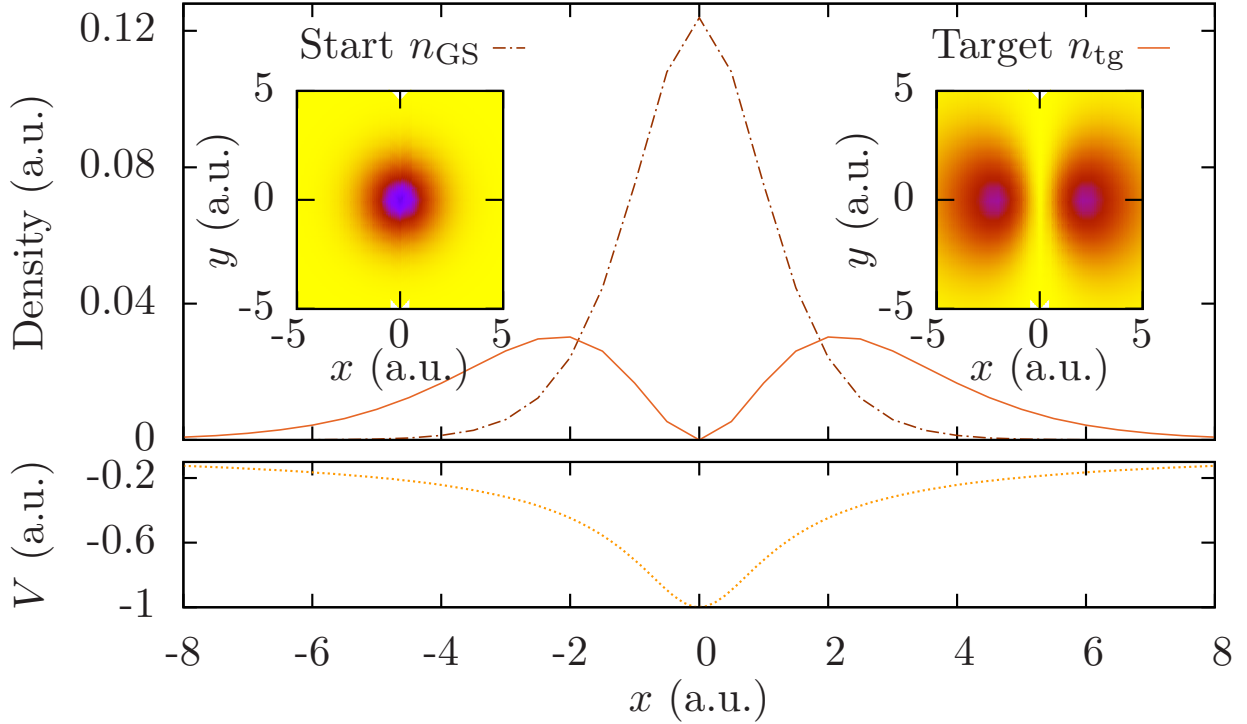} 
  \caption{(Color online) 2D H-atom. The top panel illustrates the
    densities of the $1s$ GS and the $2p_x$ ES along the $x$-axis
    ($y=0$~a.u.), their 2D plots are shown in the inset. The target state has a node
    at $x=0$~a.u. The bottom panel shows the soft-Coulomb potential $V$ along
    the $x$-axis ($y=0$).} 
    \label{fig:2d_h_stats}
 \end{center}
\end{figure}

Let us now consider a soft-Coulomb 2D model of the hydrogen atom. In
this case the objective is a stable density after a transfer from the
$1s$ GS to the $2p_x$ ES. The respective
soft-Coulomb potential reads:
\begin{align} \label{eq:hydrogen_pot}
 V(x,y) = \frac{-1}{\sqrt{1+x^2+y^2}},
\end{align}
and is shown in the bottom panel of Fig.~\ref{fig:2d_h_stats}. Such
soft potentials are used to model Wannier excitons~\cite{capaz:06}, to
describe multi-photon processes~\cite{su:91} and to understand strong
laser fields better~\cite{lein:00}. The densities of the initial and
target state $n_{\text{tg}}$ are shown in the top panel of
Fig.~\ref{fig:2d_h_stats}.

\begin{figure}
 \begin{center}
  \includegraphics[width=0.99\columnwidth,clip]{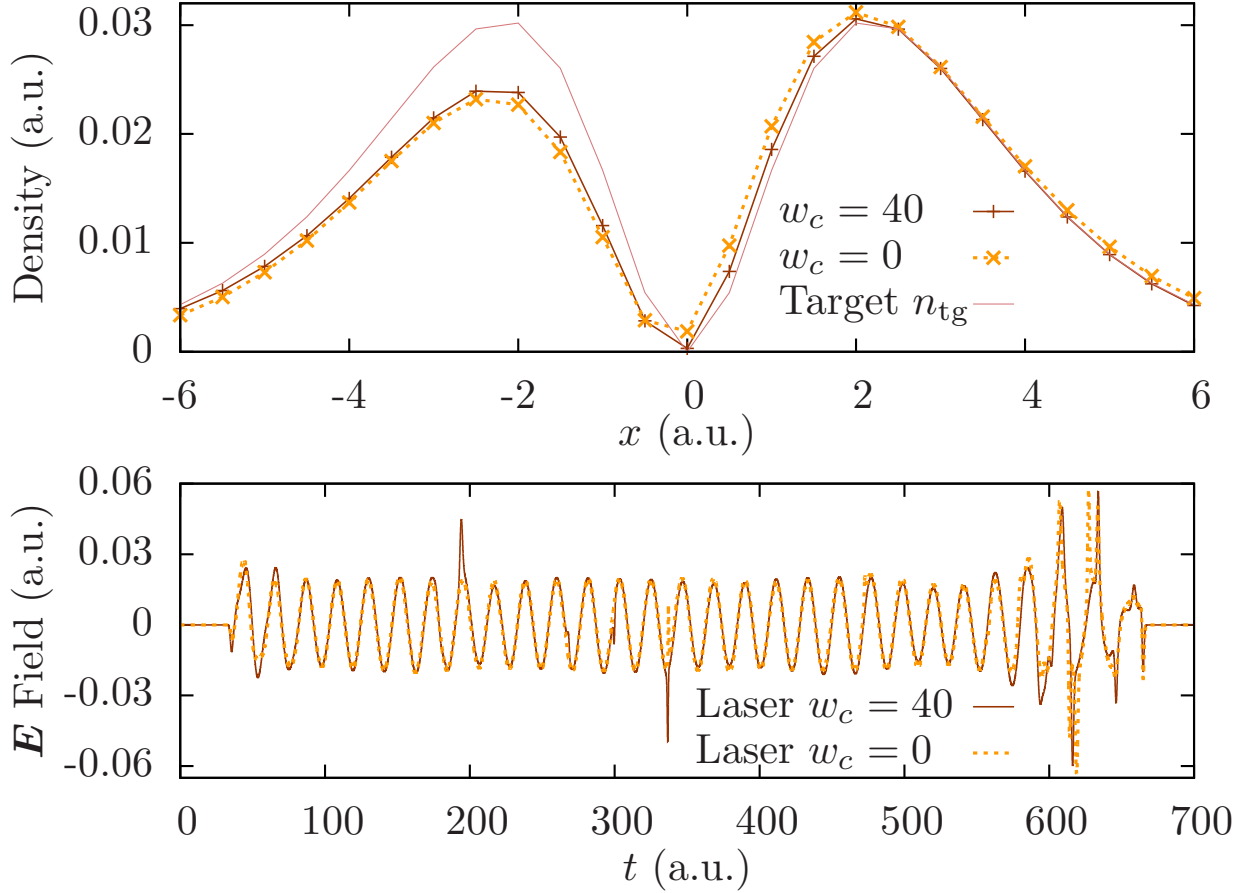} 
  \caption{(Color online) 2D H-atom. The densities at $T=700$~a.u. after an
    optimization with a control on the current ($w_c=40$) and without
    ($w_c=0$) are reported in the top panel. Only the $w_c=40$ case
    displays a good agreement with the target at the node. The bottom
    panel shows the respective optimal
    lasers. } \label{fig:2d_h_final_position}
 \end{center}
\end{figure}

At final time $T=700$~a.u. the overlap of the controlled density $n$ with the target density $n_{\text{tg}} $ is similar between the optimization with current suppression ($w_c=40$) and without ($w_c=0$) as shown in the top panel of Fig.~\ref{fig:2d_h_final_position}. It is good for $x\geq0$~a.u. (especially if current suppression enforces a solution with a node at $x=0$~a.u.), while it remains unsatisfactory for $x<0$~a.u. We underline that this asymmetry in attaining the target is a consequence of persisting density oscillations as explained below. Accordingly, the region of good overlap is periodically changing from on side of the $x$-axis to the other. Nonetheless, this asymmetry is influenced by the initial guess laser field. In fact, conserving the frequency of the guess field, but changing the sign of its amplitude leads to the optimal laser with inverted sign. This ``negative'' optimal laser causes in turn exactly the opposite terminal configuration: with a 
good overlap for $x \leq 0$~a.u. and poor overlap otherwise (not shown). Note that in 
terms of the objective functional the final density configuration and its ``mirrored'' counterpart are equally valid. The
respective optimal lasers for $w_c=40$ and $w_c=0$ are plotted in the bottom panel.

\begin{figure}
 \begin{center}
  \includegraphics[width=0.99\columnwidth,clip]{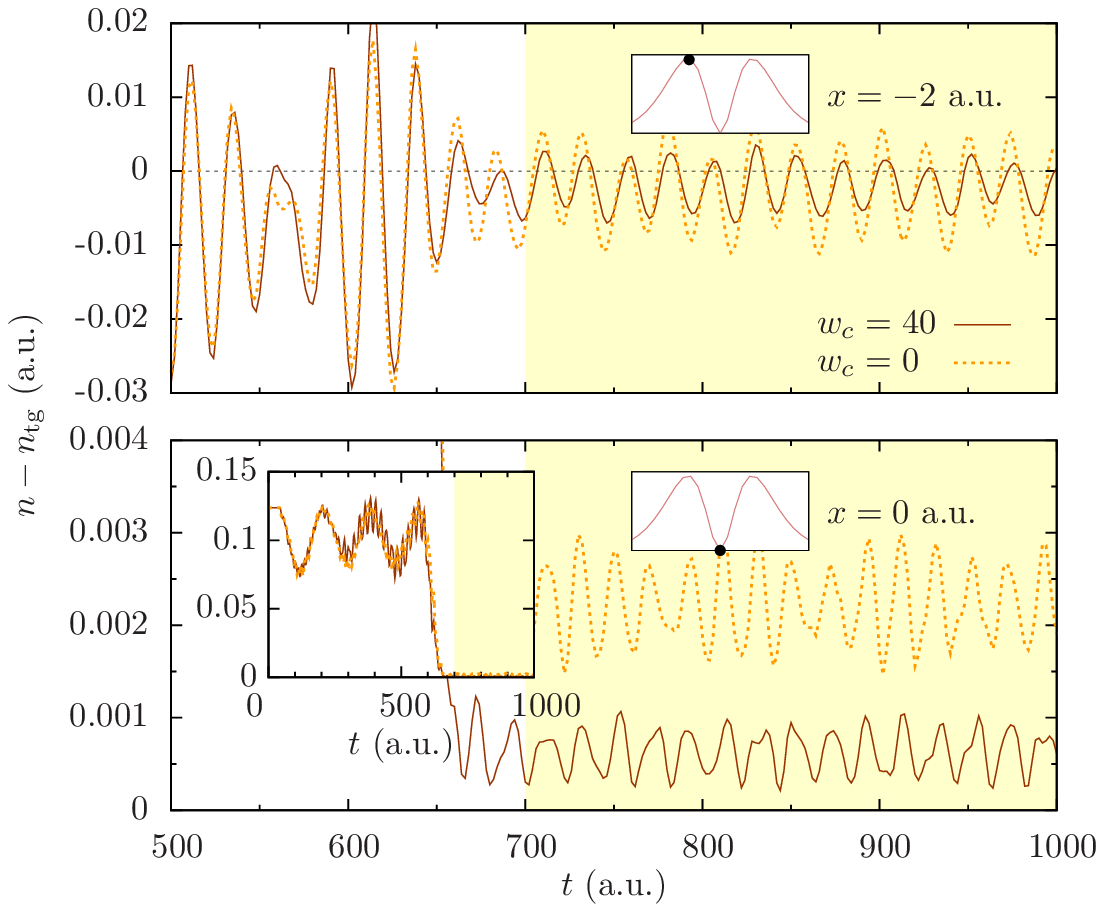}
  \caption{(Color online) 2D H-atom. The difference of the controlled
    density $n$ and the target density $n_{\text{tg}}$ as a function
    of time at a maximum ($x=-2$~a.u., top panel) and at the node ($x=0$~a.u.,
    bottom panel) of the target. Current control with $w_c=40$ is
    important because it significantly reduces the oscillations and
    decreases the difference at the node by a factor of 5. The left
    inset of the bottom panel shows the difference at the node for all
    times. The right insets in both panels are guides to the eye. They 
    show the target density with a mark on the maximum and on the node, respectively.} \label{fig:2d_h_cw40_vs_cw0}
 \end{center}
\end{figure}

However, a substantial difference between the two optimal solutions can be seen in their behavior once the control is switched off. Although we observe an oscillation of density between the two lobes in both cases, the minimization of the current leads to a significantly smaller rate of fluctuation. Figure~\ref{fig:2d_h_cw40_vs_cw0} illustrates the difference
between the controlled and target density at one maximum (at $x=-2$~a.u.,
top panel) and at the node (at $x=0$~a.u., bottom panel) of
$n_{\text{tg}}$. The difference between the target and the controlled density
at the node is reduced by a factor of five in the case of
minimized current. This means that this point of zero density is well represented if the current is held small. Consequently, only a small portion of density can pass from one side to the other. Indeed,
oscillations of the difference at the maximum are halved if the current is suppressed.

A (soft-)Coulomb potential implies the presence of degeneracies and
close-lying states, \eg the target state $2p_x$ is degenerate in
energy with $2p_y$, and a large number of energetically close states
is available. Therefore it is impossible to totally exclude a quantum
mechanical superposition of states (since the target is not a given
state, but merely a given density), and this is the origin of the
persisting oscillations in the optimal density. It is interesting to note that even our additional calculations that
involved a target in terms of the wavefunction~\cite{zhu:98b} lead to similar optimal solutions with the same problem. 
Finally, we also stress
that this kind of potential is fundamentally different from the
potentials used in the quantum well examples, since it vanishes
asymptotically (and does not grow to infinity) at large distances.
Numerically, this fact may imply the appearance of undesired border
effects, and therefore we had to use a much larger radius for the
simulation box ($r=160$~a.u.).

\section{Conclusions}
\label{sec:conclusion}

We have proposed the use of target functionals defined in terms of
both the one-particle density and the current density for QOCT
calculations. In particular, we have shown the use of functionals
that maximize the overlap of the controlled density with a given
target density, and simultaneously minimizes the current, in order to
stabilize the final state at the end of the action of the controlling
laser pulse. Such an objective fits very well into a TDDFT description
of the system. A proof of concept was offered with three prototypical
1D and 2D systems. In these cases, we observed how the suppression of
the current reduced quantum oscillations of the final state.

\section*{Acknowledgments}
D.~K. wants to thank Lauri Lehtovaara for useful discussions and
enthusiastic suggestions. A.~C. acknowledges support from the research
grant FIS2009-13364-C02-01 (MICINN, Spain). D.~K. and M.~A.~L.~M.
acknowledge support from the French ANR (ANR-08-CEXC8-008-01).
Calculations were performed at GENCI (project x2011096017).



%

\end{document}